\documentclass[12]{amsart}

\usepackage{amsmath, amssymb}
\begin{document}

\title[TQFT and K.Saito's theory of Primitive form]{TQFT, Homological Algebra and elements of K.Saito's Theory of Primitive Form: an attempt of mathematical text written by mathematical physicist}

\author{Andrey Losev}
\address{Wu Wen-Tsun Key Lab of Mathematics, Chinese Academy of Sciences\\
Scientific Research Institute of System Development;\\
National Research University Higher School of Economics;\\and Institute for Theoretical and Experimental Physics.}
 \email{aslosev2@gmail.com}






\maketitle

\begin{abstract}
The text is devoted to explanation of the concept of Topological Quantum Field Theory (TQFT), its application to homological algebra and to the relation with the theory of good section from K.Saito's theory of Primitive forms.
TQFT is explained in Dirac-Segal framework, 1 dimensional examples are explained in detail. As a first application we show how it can be used in explicit construction of reduction of $\infty$-structure after contraction of a subcomplex.
Then we explain Associativity and Commutativity equations using this language. We use these results to construct solutions to Commutativity equations and find a new proof of for the fact that tree level BCOV theory solved Oriented Associativity equations.
 \end{abstract}

\section{Preface for mathematicians: please do not be afraid to try to read this text}
Nowadays it is clear that what people are calling modern mathematical physics (some people are calling it mathematical quantum field theory) is a draft of a future chapter of pure mathematics.
 It is a very preliminary heuristic draft full of heuristic ideas and unproven conjectures. Moreover, it is a bad habit in mathematical physics texts to omit definitions and not to distinguish conjectures and theorems.
Therefore, such texts are unreadable by normal mathematicians.
These texts are mostly considered as a source of heuristic ideas that  should be properly formulated, studied and turned into mathematics. This work  of translating heuristics into mathematics is an extra skill that mathematicians may not have.
Moreover, such translation can be done only if mathematician already spent a lot of time reading this heuristics and thinking about it. Therefore, even mathematicians that have this skill prefer to save their time and wait until someone translates the mathematical physics text for them.

It is quite logical to expect that mathematical physicists would sometime write texts that are readable by mathematicians. Due to the current reputation of mathematical physicists
 the potential readability of the texts should be mentioned in the title.  Here I present an attempt to write such a text.

At the first glance the text contains many "mathematical physics wording", like quantum mechanics,  quantum field
theory, amplitudes and so on. However, I will try to define everything that I will use  keeping for these objects their standard names.

This may be compared with text on homological algebra that is using names like differential, cycles
and boundaries. In order to understand such text the reader does not need to know differential geometry.
All object are defined in terms of linear algebra, they are just keeping names from the differential geometric realization.

The text is devoted to explanation of the concept of Topological Quantum Field Theory (TQFT), its application to homological algebra and to the relation with the theory of good section from K.Saito's theory of Primitive forms.
TQFT is explained in Dirac-Segal framework, 1 dimensional examples are explained in detail. As a first application we show how it can be used in explicit construction of reduction of $\infty$-structure after contraction of a subcomplex.
Then we explain Associativity and Commutativity equations using this language. We use these results to construct solutions to Commutativity equations and find a new proof of for the fact that tree level BCOV theory solved Oriented Associativity equations.

\section{TQFT for mathematicians}
\subsection{Dirac-Segal Quantum Field Theory}

    {\bf Definition.} 
 A D-dimensional quantum field theory according to Segal \cite{Segal} may be considered as monoidal functor $I$ between monoidal category
 of decorated smooth oriented D-dimensional cobordisms  and monoidal category of vector spaces. By decorated cobordism we  mean
 that the oriented D-dimensional smooth manifold is equipped with some kind of geometrical data, such as metric, complex or almost complex structure etc.
 When manifold is cut along the $D-1$ dimensional submanifold the geometrical data is induced on the result of the cutting.

 {\bf Explanation of the definition.}
 The functor was motivated by functional integral that is why it is called  $I$.

The objects are $D-1$ oriented dimensional manifolds $\Gamma$ and $I(\Gamma)$ are vector spaces, such that if we change orientation of $\Gamma$ to $\Gamma^{opp}$ the image is the dual space:
\begin{equation}
I(\Gamma^{opp})=I(\Gamma)^*
\end{equation}

Let us call the $D$ dimensional manifolds by $\Sigma$, and geometrical data on it by $g$.
The functor sends the pair $(\Sigma,g)$ to $I(\Sigma, g)$ that for each fixed $g$ belongs to $I(\partial \Sigma)$,
so the functor $I$ may be considered as an element of the tensor product of  $I(\partial \Sigma)$ and the space of functions on the space
 $Geom_{\Sigma}$ of possible geometric data on  $\Sigma$:
\begin{equation}
I \in I(\partial \Sigma)\otimes Fun(Geom_{\Sigma})
\end{equation}

The functoriality condition means the following. Suppose that $\Sigma$ is cut by not self-intersecting and not crossing the boundaries $\Gamma$ into
$\Sigma^{\Gamma}$. The geometrical data on a cut manifold is induced from the original manifold, so we have a map:
\begin{equation}
Cut: Geom(\Sigma) \rightarrow Geom(\Sigma^{\Gamma})
\end{equation}
Its inverse image $Cut^*$ is a linear map
\begin{equation}
Cut^*: Fun( Geom(\Sigma^{\Gamma}) \rightarrow  Fun(Geom(\Sigma))
\end{equation}

 The boundary of $\Sigma^{\Gamma}$ is a disjoint union of the  boundary of $\Sigma$ and two copies of $\Gamma$ with opposite orientation, therefore
\begin{equation} \label{me}
I (\partial \Sigma^{\Gamma})=I(\partial \Sigma \cup \Gamma \cup \Gamma^{opp})=
I(\partial \Sigma) \otimes I(\Gamma) \otimes I(\Gamma)^*
\end{equation}
where we use monoidal nature of the functor that takes disjoint union into a tensor product.

Note, that there is a natural map $Glue_{\Gamma}$
 \begin{equation}
Glue_{\Gamma}: I (\partial \Sigma^{\Gamma}) \rightarrow I (\partial \Sigma)
\end{equation}
induced by a canonical contraction between $I(\Gamma)$ and $I(\Gamma)^*$.

The functoriality means that
\begin{equation} \label{cutdef}
Glue_{\Gamma} \otimes Cut^* ( I(\Sigma^{\Gamma}) ) =I(\Sigma)
\end{equation}

{\bf  One dimensional example -- Quantum Mechanics.}

If this still looks to abstract I will give an example for $D=1$. Let $\Sigma$ be an interval, and let geometric data be a metric.
Then the space of geometric data is the space of positive numbers $\mathbb{R}_+$. The boundary of an interval is a set of two points. If we denote by $V=I(point)$ for positive orientation of the point, then
\begin{equation}
I(\partial \Sigma)=I(point \cup point^{opp})=V \otimes V^*
\end{equation}
and
\begin{equation}
I(\Sigma) \in V \otimes V^* \otimes Fun(\mathbb{R}_+)=Maps(\mathbb{R}_+, End(V))
\end{equation}
Then, the equation (\ref{cutdef}) means that this map is a representation of the semigroup $\mathbb{R}_+$ in $End(V)$.

To see this, let us denote by $I_t$ the evaluation of $I$ at $t$ :
 $$I_t=ev_t I(\Sigma)$$
If we cut an interval of length $t$ by a point $P$  into two intervals with the lengths $t_1$ and $t_2$, then, due to monoidal functoriality
 \begin{equation}
I_t(\Sigma^P)=I_{t_1} \otimes I_{t_2}
\end{equation}
and the main equation (\ref{cutdef}) reads
\begin{equation}
I_{t_1+t_2}=I_{t_1} \circ I_{t_2}
\end{equation}
where $\circ$ stands for composition in $End(V)$.
So, it is the semigroup representation condition.

The universal solution to this condition is
\begin{equation}
I_t=\exp(- t H)
\end{equation}
where $H \in End(V)$ is called Hamiltonian operator in physics.

This formulation of quantum mechanics (1-dimensional QFT) was discovered by Dirac around 1930. That is why
I often call Segal formulation of QFT the Dirac-Segal formulation.

\subsection{Quantum mechanics on oriented graphs}

It is quite interesting that although Dirac-Segal formulation was originally about manifolds, the one-dimensional  quantum field theory may be generalized to include graphs.
This generalization can be made by applying the cutting axiom (\ref{cutdef}) to  oriented graphs equipped with lenghts of their edges.

{\bf Motivation for the definition.} Vertices of the graph may be decomposed into the 1-valent that we will call external and the rest that we will call internal.
 Cutting out neighborhoods of internal vertices we are cutting the graph  into intervals and "p-corollas" (by corolla we mean a  connected tree with $p$ 1-valent vertices and
$p$ edges that link these vertcies to a single $p$-valent vertices). Depending on the orientation of the graph p-corollas
should be represented as
\begin{equation}
Cor_p \in V^{ * \otimes  (p-q)} \otimes V^{  \otimes q} \otimes Fun(\mathbb{R}_+^p)
\end{equation}
Taking the limit when lengths of all edges of the corolla are tending to zero we get PROR - multi linear operation with $p-q$ inputs
and  $q$ outputs:
\begin{equation}
m_{p-q}^{q} =\lim_{t_1, \ldots, t_p \rightarrow 0} Cor_p \in  V^{ * \otimes  (p-q)} \otimes V^{  \otimes q}
\end{equation}
{\bf Definition.}
The generalization of monoidal functor associates to the oriented graph $\gamma$ a tensor-valued function on the space of lengths of edges that we call $I_{\gamma}$:
\begin{equation}
I_{\gamma}= <\otimes_{i=1}^{n_e} I(t_i) , \otimes_{\alpha=1}^{n_c} m_{\alpha}>_{\gamma} \in Fun( \mathbb{R}_+^{n_e}),
\end{equation}
where $n_e$ is the number of edges of the graph, $n_c$ - the number of internal vertices(corollas), $<,>_{\gamma}$ is the pairing determined by the incident relations of the graph.

{\bf Example 2.2.1.}

 For the Y-shaped graph with two inputs and one output the $I_Y \in V^{* \otimes 2} \otimes V$, and its value
on $v_1 \otimes v_2$ is equal to:
\begin{equation}
I_Y (v_1, v_2) =\exp(-t_3 H) ( m_2^1 ( \exp(-t_1 H) v_1, \exp(-t_2 H)v_2))
\end{equation}

{\bf Example 2.2.2.}
For a graph that is an interval consisting of 3 intervals of lengths $t_1,t_2,t_3$ the corollas are linear operators
$\phi$ and operator corresponding to such graph is
\begin{equation}
I_{3int} =\exp(-t_3 H) \phi  \exp(-t_2 H) \phi \exp(-t_1 H)
\end{equation}

{\bf Definition of preamplitude.}
Suppose that $H$ is semi positive defined. In this case it is possible to consider so-called preamplitudes $PA$ that are limits of $I_{\gamma}$ when lengths of all external edges (edges that connect external vertices to the rest of the graph) tend to
$+\infty$. In this case operators corresponding to these edges turn into projectors on the kernel of $H$ that we will
denote by $\Pi$. Thus, preamplitude corresponding to the $Y$ graph equals
\begin{equation}
PA_Y  =\Pi ( m_2^1 ( \Pi, \Pi)) \in Ker H \otimes Ker H \rightarrow Ker H
\end{equation}
and preamplitude corresponding to the 3-interval graph is a linear operator on $Ker H$ that equals to
\begin{equation}
PA_{3int} =\Pi \phi  \exp(-t_2 H) \phi \Pi
\end{equation}

{\bf Comment.} Physicists are studying amplitudes. I think it is more appropriate to introduce preamplitudes such that amplitudes are
obtained from them by integration over the space of geometrical data, see below. In particular, in string theory the measures on the moduli space are the most known examples of preamplitudes.

\subsection{Topological quantum theory in Segal formulation}

{\bf Definition of TQFT.}The notion of topological quantum field theory (TQFT) can be obtained by the following "supersymmetrization" of Dirac-Segal formulation
of quantum field theory.
The first "supersymmetrization" is the replacement the vector spaces by complexes.
We will denote differentials in these complexes by letter $Q$ that is traditional in mathematical physics literature.
The second "supersymmetrization" is the replacement of the space $Geom$ of geometrical data by a superspace
$\Pi (T) Geom$, that stands for a tangent bundle to the space $Geom$ with inversed parity of the fibers.
The space of functions on $\Pi (T) Geom$ is just the supercommutative DeRham superalgebra $\Omega^*(Geom)$.
Thus, the functor of TQFT belongs to differential form valued element of the tensor algebra of complexes:
\begin{equation}
I(\Sigma) \in V(\partial \Sigma) \otimes \Omega^*(Geom(\Sigma))
\end{equation}
Note, that supermanifold is not a set of points, so the above definition cannot be stricly speaking considered as a specialization of the definition of QFT but rather its generalization.
The functoriality axiom (\ref{cutdef}) has exactly the same form.

{\it Closeness Axom of TQFT.}
Topological QFT have an additional axiom, the axiom of total closeness of the functor.
Namely,
\begin{equation} \label{close}
(d+Q) I(\Sigma)=0
\end{equation}
where $d$ is a de Rham differential acting on $\Omega^*(Geom)$.

{\bf Comment.}
The closeness axiom means that  the differential form $ I(\Sigma)$ becomes closed
after restriction to the closed subspace in $V(\partial)$.  If we further specialize to degree zero forms in
$\Omega^*(Geom)$ it would mean that such functions are constant on connected components, i.e. they do not
depend on geometrical data, thus, they are topological invariants. This property gave the name "topological"
to such QFT, and replacement functions by differential forms and condition of being locally constant by condition
of being just closed is standard in derived mathematics.

\subsection{Universal TQFT in dimension 1}
Similarly to QFT is dimension 1, it is easy to find the universal TQFT in dimension 1.
Namely, this TQFT is governed by a complex $V$ with differential $Q$ and additional differential $G$, and
\begin{eqnarray} \label{tqm}
I(t,dt)&=&\exp (  \{ d+Q,- t G \}) = \nonumber \\
&=&  \exp(-t \{ Q, G \} - dt G)=  \exp(-t \{ Q, G \}) (1 - dt G)
\end{eqnarray}
From the first equality above it is clear that such $I$ satisfies the closeness axiom.

It follows from the (\ref{tqm}) that after restriction to function we have a 1-dimensional quantum field theory with
a very special Hamiltonian
\begin{equation}
H= \{ Q, G \}
\end{equation}
Such Hamiltonians are known as supersymmetric Hamiltonians. Such Hamiltonians were known in differential geometry.

\subsection{Examples of supersymmetric Hamiltonians}

{\bf Example 2.5.1.}
The first example is Laplacian $\Delta$ acting on the complex $\Omega^*(X)$ of differential forms on Riemannian manifold $X$ with differential $Q$ being the de Rham differential on $X$:
\begin{equation}
\Delta= \{ d, d^* \}
\end{equation}
where
\begin{equation}
G=d^*= -* d *
\end{equation}
 is the Hodge conjugated operator.

{\bf Example 2.5.2.}
The second example of such Hamiltonian acting on the same complex $\Omega^*(X)$ is the Lie derivative $Lie_v$
along the vector field $v$ on $X$:
\begin{equation}
Lie_v= \{ d,  \iota_v \}
\end{equation}
where
\begin{equation}
G=  \iota_v
\end{equation}
is the operator of contraction of differential form with the vector field $v$.

{\bf Example 2.5.3.}
The third example comes from the homological algebra.
Consider a complex $V$ together with its decomposition into a direct sum of contractable subcomplex $V_c$ and residual
subcomplex $V_r$:
\begin{equation} \label{dec}
V=V_r \oplus V_c
\end{equation}
so that we have  inclusion and projection  operators
 \begin{equation} i: V_r \rightarrow V   \; , \; \; \;  \pi  : V \rightarrow V_r
\end{equation}
and also an odd homotopy operator (that inverts differential on contractable subcomplex $V_c$)
 \begin{equation} h: V \rightarrow V   \; , \; \;
   h\circ h=  h \circ i = \pi  \circ h =0  , \; \; \{Q,h \}=Proj_{V_c}=id -  i \circ \pi,
\end{equation}
here $Proj_{V_c}$ stands for the projection operator on a contractable subcomplex.

This data means that we have a topological quantum mechanics with
\begin{equation}
G= h,  \;\;\;  H=Proj_{V_c}
\end{equation}

The $I$ of this example has another important property that we will use below, namely, it has a limit when $t$ turns to
$+\infty$:
\begin{equation}
\lim_{t \rightarrow +\infty} \exp(-tH)= \pi \circ i=Proj_{V_r},
\end{equation}
where $Proj_{V_r}$ is the projector on the space $V_r$.

\subsection{Local observables in QFT}
By Segal's definition the main object $I$ in QFT corresponds to manifolds with boundaries.
Now we define the notion of observables in QFT.
Consider the following construction: take a submanifold $C$ in $\Sigma$ and consider its small tubular neighborhood $Tub \; C_{\epsilon}$, i.e. the space of points in $\Sigma$ such that their distance to $C$ is not greater than $\epsilon$.
Let $\Gamma_{C,\epsilon}$ be the boundary of $Tub \; C_{\epsilon}$.
Consider the space $\Sigma_{Tub \; C_{\epsilon}}$ obtained by cutting the tubular neighborhood $Tub \; C_{\epsilon}$ out of
$\Sigma$:
\begin{equation}
\Sigma_{Tub \; C_{\epsilon}}=\Sigma  \backslash  Tub \; C_{\epsilon}
\end{equation}
The boundary of $\Sigma_{Tub \; C_{\epsilon}}$ is a disjoint union of the boundary of  $\Sigma$ and $\Gamma_{C,\epsilon}$.
Therefore
\begin{equation}
I_{\Sigma_{Tub \; C_{\epsilon}}} \in V_{\partial \Sigma} \otimes V_{\Gamma_{C,\epsilon}}
\end{equation}

{\bf Definition of observables.}
We will call the family of vectors $v_{\epsilon} \in V^*_{\Gamma_{C,\epsilon}}$ good if  there is a limit
$$
\lim_{\epsilon \rightarrow 0} < v_{\epsilon} , I_{\Sigma_{Tub \; C_{\epsilon}}}>
$$
The good family of vectors is called null if this limit equals to zero.
We define the space of observables $Obs_C$ associated to $C$ as the coset of good families
over the space of null families.
The limit mentioned above is defined on the space of observables, and we will call the value of this limit the correlator
of an observable associated with the submanifold $C$:
\begin{equation}
I(O_C) = \lim_{\epsilon \rightarrow 0} < v_{\epsilon} , I_{\Sigma_{Tub \; C_{\epsilon}}}>
\end{equation}

If $C$ is point we will call such observables local observables.

{\bf Example.}
Consider the one-dimensional example. The only observables are local observables. The tubular neighborhood of a point on an interval is also an interval, so local observables are just linear operators:
$$
Obs_{P} = End (V)
$$
and correlator of an observable in topological quantum mechanics equals to
\begin{equation}
I(O_P)(t_1,t_2)=\exp (-t_1 H - dt_1 G) O \exp (-t_2 H - dt_2 G)
\end{equation}
where $t_1$ and $t_2$ are the distances between the point $P$ and the left and right ends of the interval respectively.
We see that in 1-dimensional theory the observable corresponds to the 2-valent internal vertex.

{\bf Remark. Local observables as tangent vectors to space of theories.}
It is important to note that local observables correspond to tangent space to the space deformations of QFT that do not change the vector spaces associated to boundaries. It is a general statement, but we may illustrate it in  topological quantum mechanics.
Actually, if we change $Q$ by $\delta Q$, the $I$ corresponding to the interval of length $t$ is changing according to:
\begin{equation}
\delta I=
 \int_{t_1+t_2=t, t_i \geq 0}
\exp (-t_1 H - dt_1 G) \delta Q \exp (-t_2 H - dt_2 G)
\end{equation}

\section{Topological quantum mechanics on trees and constructive induction of operations after contraction of the subcomplex}

Here I will explain how topological quantum mechanics (another name of 1-dimensional TQFT) of Example 2.5.3 on a tree may be used to give a constructive proof of the Kadeishvili-type theorems \cite{Kad} about induction of $\infty$-structures on residual subcomplexes ( like subcomplex $V_r$ in (\ref{dec})).

\subsection{Amplitudes and Kadeishvili theorem.}

Consider  preamplitudes in topological quantum mechanics on rooted oriented  trees.
To define such theory we need a triple consisting of $Q,G$,  and a set of operations $m_1, \ldots, m_k$ ,where $Q$ and $G$ will be as in Example 2.5.3 and $m_n \in V^{* \otimes n}\otimes V$.

 In the case from example 2.5.3 the projector $\Pi$ associated to leaves may be considered as operator $i$ of inclusion of $V_r$ into $V$, and projector $\Pi$ associated to the root may be considered as a projection $\pi$ from $V$ to $V_r$.
Let $n_l$ be the number of leaves of the rooted tree and $n_c$ the number of internal vertexes (corollas) . The preamplitude associated to tree $\gamma$ is a $n_l$-operation on $V_r$  taking value in differential forms on the lengths of edges. It has the following form:
\begin{equation}
PA_{\gamma}= <\otimes_{i=1}^{n_e} I(t_i) , \otimes_{\alpha=1}^{n_c} m_{\alpha}\otimes i^{\otimes n_l} \otimes \pi >_{\gamma} \in \Omega^*( \mathbb{R}_+^{n_e}) \otimes  V_r^{* \otimes n_l} \otimes V_r ,
\end{equation}
where contraction of tensors are made according to the tree.

For example, for the $Y$ tree the preamplitude is just the restriction of the binary operation on $V_r$, while for
the 3-interval tree ( that tree has only $m_1$ operation $\phi$), the preamplitude equals
\begin{equation}
PA_{3int}=\pi \phi \exp(-t Proj_{V_c}-dt h) \phi i
\end{equation}

{\bf Definition of a tree amplitude.}
Let as define a tree amplitude $A_{n_l} \in  V_r^{* \otimes n_l} \otimes V_r$ as a sum over all rooted trees with
$n_l$ leaves of the integrals of preamplitudes  over spaces of lengths of edges (and we will divide each integral over the number $n_{\gamma}$ of elements of the symmetry group of the tree):
\begin{equation}
A_{n_l}=\sum_{\gamma} 1/n_{\gamma} \int_{\mathbb{R}_+^{n_e}} PA
\end{equation}
Such integral can be easily calculated, since
\begin{equation}
\int_{\mathbb{R}_{+}} \exp(-t Proj_{V_c}-dt h)=-h,
\end{equation}
and we are coming to the following formula for the amplitude:
\begin{equation}
A_{n_l}= \sum_{\gamma} (-1)^{n_e}/n_{\gamma}  <h^{ \otimes n_e} ,  \otimes_{\alpha=1}^{n_c} m_{\alpha}\otimes i^{\otimes n_l} \otimes \pi >_{\gamma} ,
\end{equation}

{\bf One leaf example.}
In particular, if the only operation is $m_1$ (that we still denote by $\pi$) the amplitude for one leaf equals to
\begin{equation} \label{intfor}
A_1=\pi \phi i - \pi \phi h \phi i + \pi \phi h \phi h \phi i - \ldots
\end{equation}

{\bf Kadeishvili Theorem.}

Suppose that operations $Q+m_1, m_2, m_3 \ldots $ form an $L_{\infty}$ algebra, then operations constructed from $Q$ and amplitudes $Q+A_1, A_2,A_3 \ldots $ also form an  $L_{\infty}$ algebra. This algebra is called the algebra obtained by the contraction of the subcomplex.

\subsection{Kadeishvili theorem in one leaf case.}

Before we proceed to discussion of the proof it is instructive to look at the particular case, namely, the case where the only nonzero operation is $m_1=\phi$.
Then the theorem actually states that if $\phi$ is a solution to Maurer-Cartan equation
\begin{equation} \label{mc1}
\{ Q, \phi \} + \phi^2=0
\end{equation}
then
\begin{equation} \label{ire}
\{ Q, A_1 \}+A_1^2=0
\end{equation}

{\bf Remark.}
We may specialize even further, and consider $Q$-closed $\phi$ and take $V_r$ to be cohomology of $Q$, so
 the input of the construction is a bicomplex. Terms in  (\ref{intfor}) look like differentials in spectral sequence construction. However, there is an important difference. In spectral sequence construction differentials act on different spaces, each
of these spaces are cohomology of the previous differential, while all summands of   (\ref{intfor}) act on the same space,
namely, cohomology of $Q$ (the first differential of the spectral sequence).
Moreover, differentials in spectral sequence construction are defined canonically while operators in  (\ref{intfor}) depend on the choice of the direct decomposition of the space $V$ into cohomology and contractable complex , and also on the choice of homotopy.
So, differential $A_1$ seem to contain more information than differentials of spectral sequence, however, one can show that all information that is invariant under the choice of decomposition and homotopy is contained in spectral sequence differentials. On the other hand, construction  (\ref{intfor}) provides an explicit answer to the question: how to construct a differential on the cohomology of $Q$, such that cohomology of the constructed differential would be equal to
cohomology of the total differential ( if the sum in  (\ref{intfor}) contains finite number of terms). We will later see the similar phenomena in discussion of Massey operations.

Now let us discuss how to prove a theorem.

\subsection{Outline of the  direct proof of Kadeishvili theorem}

The subtle issue in the one leaf case is that $\phi$ has grading zero, so the sum (\ref{intfor}) may diverge. Thus,  we have to consider it as a formal one,  namely, we take
\begin{equation}
\phi=\sum_{k=1}^{\infty} \phi_k \epsilon^k
\end{equation}
and consider $\epsilon$ as formal parameter.
From Maurer-Cartan equation we get
\begin{equation} \label{mcg}
\{ Q, \phi_1 \}=0, \; \; \{ Q, \phi_2 \}=-\phi_1^2,  \; \; \ldots
\end{equation}

In the first order in $\epsilon$ statement of the theorem is Q-closeness of $\pi \phi_1 i$ that follows from Q-closeness of
$\phi_1$. In the second order the substitution of expression  (\ref{intfor}) into  (\ref{ire}) gives
\begin{eqnarray}
&& \{Q,  \pi \phi_2 i -\pi \phi_1 h \phi_1i \}+\pi \phi_1 i \pi \phi_1 i = \nonumber \\
&& \pi \{Q, \phi_2 \} i +\pi \phi_1 \{Q, h \} \phi_1 i + \pi \phi_1 (1- \{Q, h \}) \phi_1 i=0
\end{eqnarray}
where we used (\ref{mcg})

One may show that similar cancellations happen at higher orders.

The multileaf case can be treated similarly, by direct computation and combinatorics of trees. The composition of amplitudes would produce $i \pi$ term that can be replaced by $ (1- \{Q, h \})$.
The contribution coming from $1$ being combined with the action of $Q$ on the operation $m_n$ would lead to the
Maurer-Cartan expressions, while $ \{Q, h \}$ terms would be cancelled by the action of $Q$ on homotopy.
This may be considered as a combinatorics trick. It is reasonable to look for more conceptual explanation, that we will give in the next subsection.
It is remarkable that the more conceptual explanation has generalizations in dimensions higher than 1.

\subsection{ Proof of the induction theorem based on the properties of the preamplitude}

In order to explain main ideas we will first give the proof in the simplest case when $V_r$ is the space of cohomology of $Q$ and when both terms in Maurer-Carttan equation are equal to zero separately, i.e. when all operations $m_k$ are
$Q$-closed and when these operation form the infinity algebra. Later we will comment on how this proof may be modified  to include the general case.

The proof that we are going to present here is based on two properties: closeness and factorization.
Closeness means that each preamplitude is closed as a form on the space of lengths of edges.
It follows from $d+Q$ closeness of the evolution operator $I(t,dt)$ associated to intervals and from $Q$-closeness of
operations $m_i$, projection $\pi$ and inclusion $i$.

 Factorization means that when length of one of the edges goes to infinity, the preamplitude factorizes into composition
of preamplitudes corresponding to trees obtained by cutting the original tree along such an edge.
This follows from
\begin{equation}
\lim_{t \rightarrow +\infty} \exp(-t Proj V_c)(1-dt h) =Proj V_r=i \circ \pi
\end{equation}

The idea of the proof is to compactify the space of lengths of the tree to a $n_e$ hypercube, to take a $n_e-1$-form component
of the preamplitude and to integrate it along the boundary of the hypercube.  Since the preamplitude is closed its  $n_e-1$-form component is also closed. Thus, the integral will be equal to zero. Contributions from the faces corresponding to infinite lengths trees would form compositions of amplitudes. Contributions from zero length faces
would form an amplitude corresponding to the tree that has a composition of operations in one of its internal vertices.
When we sum over all trees contribution of such vertices would cancel due to infinity algebra conditions on the operations. For the one leaf trees this derivation was obtained by Lysov \cite{Lysov}.

It is important to mention that the space of lengths for the one leaf case may be considered as a moduli space $RM_n$:
for points on a real line moduli common shifts:
\begin{equation}
RM_n=\mathbb{R}^n/ \mathbb{R}
\end{equation}
In the next section we will study the complex version of such moduli space.



{\bf Remark 1.}
This idea was used by Kontsevitch\cite{KonDQ} in his proof of $\infty$-morphism theorem in the work on deformation quantization.
He used the moduli space of complex structures on a disk with marked points in the bulk of the disk and on the boundary of the disk. The two-dimensional topological theory he constructed produced closed differential form on the moduli space.
The boundary contribution turn out to be exactly the $\infty$-morphism statement.

{\bf Remark 2.} If $V_r$ equals to cohomology of $Q$ then the amplitudes for the multileaf case may  be considered as generalizations of Massey operations in the same way in which $A_1$ is the generalization of the spectral sequence differentials. While Massey operations are conventionally understood as a partly defined canonical operations (when some products in cohomology are zero) the amplitudes are always defined but depend on the choice of decomposition of the complex and the choice of homotopy. The amplitudes form an $L_{\infty}$ structure, and changes of auxillary data just change this structure into an equivalent one.

\section{Configuration spaces in two dimensional theories}
In the previous section we observed that it is useful to consider preamplitudes in one dimensional
quantum field theories. These preamplitudes are closed forms on moduli spaces (that sometimes are configuration spaces). Integrals of preamplitudes over moduli spaces turn out to satisfy interesting quadratic equations. In one-dimensional case amplitudes turn out to be induced operations on cohomology, while quadratic equations become infinity structure equations.

Now we would like to play similar game for 2-dimensional conformal field theories.

\subsection{WDVV equations}
Consider Deligne-Mumford compactification $\bar{M}_{0,n+1}$ of the moduli space of $\mathbb{CP}^1$ with
$n+1$ marked points. Let us take $n$ points with one orientation (we will call these points input points) and one point with the opposite orientation (we will call this point an output point).
Due to general properties of TQFT explained in the previous sections, the preamplitude in topological theory
associates vector spaces $W$ to input points, vector space $W^*$ to an output point (where $W$ is the space of cohomology of $Q$ in the space of local observables).QFT associates to $\mathbb{CP}^1$ with $n$ marked points a preamplitude
\begin{equation}
PA_n \in \Omega^*(\bar{M}_{0,n+1}) \otimes Hom( W^{\otimes n}, W)
\end{equation}
This preamplitude satisfies two main properties:\\
1.Symmetry under permutations of  inputs\\
2.Closeness, i.e.
\begin{equation}
d PA_n =0
\end{equation}
where $d$ is De Rham differential on $\bar{M}_{0,n+1}$.\\
3.Factorization. Consider the compactification divisor $D_{n_1,n_2}$ of the moduli space $\bar{M}_{0,n+1}$ where projective space  degenerates into two projective spaces with $n_1+1$ on the component that does not contain an output
point and with $n_2+1$ points on a component that has an output point.
Being restricted to such a divisor the preamplitude factorizes:
\begin{equation}
PA_n |_{D_{n_1,n_2}}=PA_{n_2} \circ PA_{n_1}
\end{equation}

Therefore we can get quadratic relations on amplitudes by integrating preamplitude against boundary divisors that equal to zero in homology of moduli space.

The simplest case is provided by $\bar{M}_{0,4}$ where we will label input points as 1,2 and 3. This moduli space is just
$CP^1$. Compactification divisor represents surfaces that are two projective spaces intersecting by a point with two input points belonging to first $\mathbb{CP}^1$ and the third input point and the output point belonging to the second $\mathbb{CP}^1$. Since the moduli space is
connected the difference of divisors ( that we will call $D_a$) equals to zero in homology of moduli space. We call it $D_a$ because evaluation of amplitudes on this divisor leads to associativity of two to one amplitudes.

This relation leads to relations in other moduli spaces (known as Keel's relations \cite{Keel}). Consider a forgetful map
\begin{equation} f: \bar{M}_{0,4+k} \rightarrow \bar{M}_{0,4}
\end{equation}
The divisors  $f^*(D_a)$ are  compactification divisors. They correspond to surfaces that are  two copies of $\mathbb{CP}^1$  intersecting at a point with input points spread among components (such that there are at least two input points
on a component that does not contain an output point, and at least one input point on the component that has an output
point).

Consider the total amplitude TA  that is a sum over all amplitudes and, therefore, the map
\begin{equation}
TA   \in \oplus_{n=2}^{+\infty} S^n W \rightarrow W,
\end{equation}
as a formal vector field $v$ on $W$.

In terms of the vector field $v$ the Keel's relations take the following form.
Let $T^a, a=1,\ldots , \mu$ be linear coordinates on $W$.
Consider second partial derivatives   $f_{bc}^a$ :
\begin{equation}
f_{bc}^a = \frac{\partial^2 v^a}{\partial T^b \partial T^c}
\end{equation}
The Keel's relation state that $f_{bc}^a$ form structure constants of associative commutative algebra - the Oriented Associativity equation:
\begin{equation} \label{oae}
\sum_{a=1}^\mu f_{bc}^a f_{ad}^e =  \sum_{a=1}^\mu f_{cd}^a  f_{ba}^e
\end{equation}

Suppose, that the space $W$ has a non degenerate scalar product $\eta$, such that after identification of $W$ and $W^*$ the $n$ to one amplitudes become symmetric in all $n+1$ arguments. In this case the vector field has a
potential form, namely
\begin{equation}
v^a=\sum_{b=1}^{\mu} \eta^{ab} \frac{\partial F}{\partial T^b}
\end{equation}
The Oriented Associativity equations for potential vector field are known as WDVV or simply Associativity equations.

\subsection{Losev-Manin moduli spaces, simplest space and Commutativity equations}

In \cite{LoMa} Losev and Manin introduced a new moduli space for Riemann surfaces with marked points.

{\bf Definition of Losev-Manin moduli spaces.}
The marked points are colored into to colors - black and white. When a handle degenerates or a surface is decoupled into two surfaces glued by a point the degeneration is described by white points (one input and one output). White points behave exactly like marked points in Deligne-Mumford compactification of the moduli space. Black points are distinguished from the white points
because they can collide with other black points. However, black points cannot collide with white points.

In construction of the amplitudes vector spaces associated to black points are different from that associated to white
points - I will denote these spaces by $V$.

The simplest example of Losev-Manin moduli space is the moduli space $L_k$ of projective space with two white points (one input and one output) and
$k$ black points (for stability $k$ should be positive) .

The space $L_k$ may be described as a compactification of a coset $(\mathbb{C}^*)^{k}/\mathbb{C}^*$, where $\mathbb{C}^*$ is acting diagonally
by multiplication. Really, input white point may be placed at zero, output point - at infinity, $(\mathbb{C}^*)^{k}$ is the space of positions of black points and $\mathbb{C}^*$ is the action of automorphism of the $\mathbb{CP}^1$ with two marked white points.

The non-compact directions in complex codimension 1 are formed when a group of $k_1$ points tends to zero (or, equivalently, a group of $k-k_1$ points tends to infinity. Such limits are compactified by two $\mathbb{CP}^1$ such that infinity
of the first $\mathbb{CP}^1$ is identified with the zero of the second $\mathbb{CP}^1$. First $\mathbb{CP}^1$ carries $k_1$ black points while
second $\mathbb{CP}^1$ carries $k-k_1$ black points. We will call such divisors $D_{k_1,k_2}$.

The moduli space restricted to divisor $D_{k_1,k_2}$  is $L_{k_1} \times L_{k_2}$.

The preamplitudes are closed differential forms on $L_k$ with values in $S^k V^* \otimes End(W)$, and similarly to the Deligne-Mumford case we should have a factorization condition:
\begin{equation}
PA_k |_{D_{k_1,k_2}}=PA_{k_2} \circ PA_{k_1}
\end{equation}
where composition is a composition in $End(W)$.

Like in the case of Keel's relations there is a fundamental compactification divisor that equals to zero in homology.
Consider $L_2$, it may be parametrized by ratio of coordinates of two black points: $w=z_1/z_2$.  Since $L_2$ is connected the divisor
$$
D_c=\{w=0 \}-\{w=\infty \}
$$
is zero in homology of $L_2$.
This mean that if we consider $PA_1$ as a map from $V$ to $End(W)$ then
\begin{equation}
[ PA_1(u_1), PA_1(u_2)]=0
\end{equation}
Due to this commutativity we call divisor $D_c$ commutative divisor.

 Like in the Deligne-Mumford case there are forgetful maps
\begin{equation}
 f: L_{k+2} \rightarrow L_2
\end{equation}

Therefore, preimages of commutative divisor $f^*(D_c)$ also equal to zero in homology.
It turns out that these equations may be written as follows. Let us consider $A_n$ as a n-th coefficient of Teylor seria
of the $End(W)$ valued function $A_{tot}$ on $V$. Let us denote as $d^v$ the De Rham operator on $V$. Then equations, that we would call  Commutativity equations, look
as follows:
\begin{equation}
d^v A_{tot}\circ d^v A_{tot}=0,
\end{equation}
where composition implies simultaneous composition as elements of $End(W)$ and external multiplication as 1-forms on $V$.
Quite remarkably, this equation first appeared in the Theory of Primitive form of K.Saito.
\subsection{Construction of solutions to Commutativity equation from bicomplexes with strong Hodge property}
In this subsection I will explicitly construct preamplitudes on spaces $L_k$ in terms of linear algebra data.
I will give explicite construction in simplest case and briefly explain how it may be generalized to more general case.
Moreover, I will concentrate on linear data with strong Hodge condition, for generalization to Hodge condition in terms of
Khoroshkin, Markarian and Shadrin, see their paper \cite{KMS}.

{\bf Definition of strong Hodge property.}
Consider $Z_2$ graded bicomplex $V$ with differentials $Q$ and $G_-$. Let $W$ be cohomology of $Q$, $G$-homotopy,
and let $i_W$ and $\pi_W$ be inclusion of cohomology and projection to cohomology respectively.
  By strong Hodge property I mean the following conditions:
\begin{equation}
G_- i_W=\pi_W G_{-}=0 \; \; \; \{ G, G_{-} \}=0
\end{equation}

In order to construct solution to Commutativity equation we need a commutative family.

{\bf Definition of simplified commutative family.}
A simplified commutative family is an even formal map $U$ from $V$ to commutative subalgebra of $End(W)$.
We demand that this map
has the simplified Maurer-Cartan property:
\begin{equation} \label{smc}
[Q,U]=0;  \; \; \; [ [G_{-}, U],U] =0
\end{equation}

{\bf Construction of preamplitude for $L_k$ spaces.}

We construct a differential form on $L_k$ as follows.
Let us parametrize points on $\mathbb{C}^k$ by their coordinates
$$
z_a=\exp(t_a+i \phi_a)
$$
and denote
$$
t_{aa-1}=t_a-t_{a-1}, \; \; \; \phi_{aa-1}=\phi_a-\phi_{a-1}
$$
Then the preamplitude $PA_k$ has the following form:
\begin{eqnarray}
PA_k & = & \pi_W U \exp(- t_{kk-1} \{ G, Q\} -dt_{kk-1} G +i d\phi_{kk-1} G_{-})U \ldots
\nonumber \\
&&  \ldots U \exp(- t_{21} \{ G, Q\} -dt_{21} G +i d\phi_{21} G_{-})U i_W
\end{eqnarray}

Now consider $d^V PA_k$. It follows from the properties described above that $d^V PA_k$ is closed differential form on the space $L_k$.Actually, we need to check that this form has no jumps when $t_{aa-1}$ reaches zero, and that
it can be continued to compactification. The second property follows from the fact that $G_{-}$ annihilates cohomology $W$. The first property  is more tricky, it follows from commutativity of the algebra and from
\begin{equation}\label{aax}
\{ d^VU, [G_{-}, U] \}=0
\end{equation}
that can be derived from the simplified Maurer-Cartan conditions (together with the commutativity of the target algebra).
Factorization property of $PA_k$ is obvious, therefore by integrating over spaces $L_k$ and summing over $k$ we obtain that
\begin{equation} \label{cef}
A= d^V (\sum_{k=0}^{+\infty} \pi_W (U (- G G_{-}U)^k) i_W)
\end{equation}
squares to zero, i.e. $A$ solves Commutativity equation.
In the proof presented above it is clear that $G_{-}$ is an operator of contraction with vector field of the circle action,
and K.Saito differential (see last section):
\begin{equation}
Q_S=Q+zG_{-}
\end{equation}
is an equivariant differential with respect to the circle action.

{\bf Purely homological proof.}

It is also instructive to give another proof of the formula (\ref{cef}), that belongs purely to standard homological algebra.
Consider the  bicomplex  $V\otimes \Omega^*(V)$, where the first differential equals to $Q$ and the second differential
$\phi$ is given by
 \begin{equation}
\phi= d^VU + [G_{-}, U]
\end{equation}
The $\phi$ squares to zero due to (\ref{aax}).

Now we induce the action of $\phi$ on cohomology of $Q$, i.e. on $W$ according to (\ref{intfor}).
We get $A_1$ that contains differential forms on $V$ of different degrees. Since $G_{-}$ annihilates $W$ the degree zero form is absent. The relation $A_1^2=0$ starts with equations on one forms, that we will denote a simply $A$.
From explicit form of $\phi$ we deduce that $A$ actually has the form as in (\ref{cef}), that completes the proof.

Despite the simplicity of this proof, it hides the two-dimensional origin of the formulas.

Both proofs may be generalized to the non simplified Maurer-Cartan case, i.e. that is
\begin{equation} \label{nsmc}
[ Q,U]+[[G_{-}, U],U]=0
\end{equation}

\subsection{Tree level BCOV theory as a solution to Oriented Associativity equations}

In the previous section we constructed solution to Commutativity equation in terms of linear algebra.
Recall, that moduli space $\bar{M}_{0,k}$ may be obtained from spaces $L_{k-2}$ by blowing up diagonal in the configuration space of black points. Therefore, one may conjecture that construction similar to construction of
solution to Commutativity equation may be obtained by replacing line with points by a trivalent trees.

Another argument that supports this idea is the concept of tropicalization of Riemann surfaces, that replaces general surfaces by trivalent graphs. One may think that cohomology of the moduli space of tropical surfaces  are equal to cohomology of $\bar{M}_{0,k}$.

Based on this heuristic arguments we will construct the solution to Associativity equations in terms of algebraic data.
Such construction was known as BCOV theory, however, it was only in the work of Losev and Shadrin \cite{LoSha} when it was
proven that this construction actually gives solution to Associativity equations. Proof in that paper was combinatorics and quite involved. Here we will outline the simple proof based on the construction of solution to Commutativity equation.

We will explain here oriented version, however, all arguments may be generalized to non oriented one.

{\bf Construction of the solution to Oriented Associativity equation.}

Suppose we are given the following data:\\
1. A supercommutative associative differential $Z_2$-graded algebra $B$ with multiplication $m$ and differential $Q$.\\
2. A second order odd differential operator $G_{-}$ that squares to zero, anticommutes with $Q$ and annihilates $Q$-cohomology.\\
3. A homotopy $G$ that contracts an algebra to its cohomology $W$ and that anticommutes with $G_{-}$.\\

We construct the following operation $A_{\gamma}$ for a rooted oriented tree $\gamma$ with $n_l$ leaves  :
\begin{equation}
A_{\gamma}: W^{\otimes n_l} \rightarrow W,
\end{equation}
just like in the theory of induced operations, with the only change - edges correspond not just to homotopy but to
a product $G G_{-}$ like in construction of solution to Commutativity equations.

{\bf Losev-Shadrin theorem:}
\begin{equation} \label{bcov}
A=\sum_{\gamma}  A_{\gamma}/n_{\gamma},
\end{equation}
where the $n_{\gamma}$ is the order of the groop of symmetry of the graph, is a formal vector field on $V$ that satisfies Oriented Associativity equation (\ref{oae}).

{\bf Outline of the new proof that uses Commutativity equation:}\\
1. Pick up a leaf of graph and connect it to the root. We get a line with various trees attached to it at points,
that are acting on the line as linear operators of $End(B)$ (roots of trees are inserted at one of the entrances of multiplication operator). Sum over all trees and consider it as an element $X$ of $End(W)$.\\
2.Check that these operators (after summing up trees) satisfy Maurer-Cartan equation.\\
3. Therefore, the element $X$ satisfies Commutativity equation and also is symmetric under interchange of two $W$ -
one that corresponds to distinguished leaf and a typical leaf of an original tree. Thus, it is a solution to Oriented
Associativity equation.\\

{\bf Remark.} The expression (\ref{bcov}) was defined as an amplitude in the BCOV theory \cite{bcov}. However,
authors did not pay attention to the fact that it actually solves Associativity equations.

\section{Good sections and Commutativity equation in K.Saito theory of Primitive form}

Consider $W$ - the   quasihomogenious polynomial of $n$ variables $X_1, \dots , X_N$. Consider the ideal $I$ generated by partial derivatives $\frac{\partial W}{\partial X_i}$ and the Milnor ring:
\begin{equation}
R=\mathbb{C}[X_1, \ldots , X_N] /I
\end{equation}
Assume that the Milnor ring is finite dimensional and its dimension as a vector space over complex number is $\mu$.
Consider a basis in this ring and pick up polynomials $\Phi_k$ that represent classes of $I$. Then we can form a family
\begin{equation}
W_t=W+\sum_{k=1}^{\mu} t_k \Phi_k
\end{equation}
This family is called a versal deformation of the singularity. In what follows we will consider it only over a formal disk in $t$ variables, i.e. over the $Spec \; \mathbb{C}[[t_1, \ldots, t_{\mu}]]$.

For example, for $N=1$, $W=X^n$ (we omit the subscript $1$ in the one-dimensional case), the $\mu=n-1$. If we take the lowest order representatives of the elements of the Milnor ring we will get
\begin{equation}
W_t=X^n+\sum_{k=1}^{n-1} t_k X^{k-1}
\end{equation}

Consider the de Rham complex $\Omega$ of polynomial differential forms on $\mathbb{C}^N$ with differential $d$.  Define
\begin{equation}
\Omega_t=\Omega \otimes \mathbb{C}[[t_1, \ldots, t_{\mu}]]
\end{equation}
We consider $\Omega_t$  as a complex with the same differential $d$ - that is just de Rham differential in $X$ direction.

All above was kind of standard and well-known. However, based on the study of periods of the hypersurafces $$W_t=0$$ and later on the study of exponential integrals of the form $$\int \exp(W/z)\omega $$ K.Saito  \cite{Sa} extended the complex to
\begin{equation}
\Omega_{t,z}=\Omega_t \otimes \mathbb{C}[[z]]
\end{equation}
and introduced the differential
\begin{equation}
d_{t,z}^S=z d+dW_t
\end{equation}

Let $H_{t,z}$ denote the cohomology of $d_{t,z}^S$ in $\Omega_{t,z}$.

Multiplication by $\Phi_k$ commutes with $d_{t,0}^S$, therefore, it acts on cohomology
$H_{t,0}$ by a linear operator that we will denote by $C_k$:
\begin{equation}
C_i  : H_{t,0}     \rightarrow H_{t,0}
\end{equation}
This operator will be very important in what follows.

The cohomology   $H_{t,z}$ form a vector bundle over  $Spec \; \mathbb{C}[[t_1, \ldots, t_{\mu}]]$, but this bundle does not have a natural flat connection.

However, if we localize at $z=0$, i.e. consider
\begin{equation}
\hat{\Omega}_{t,z}=\Omega_t \otimes \mathbb{C}<<z>>
\end{equation}
and corresponding cohomology $\hat{H}_{t,z}$, there is a canonical flat connection called  Gauss-Manin connection by K.Saito. This connection is given by its covariantly flat sections
\begin{equation}
S^{GM}=  [  \exp( - \frac{1}{z} \sum_{k=1}^{\mu} t_k \Phi_k) \omega]_{d_{t,z}^S}
\end{equation}
where $\omega$ is $t$-independent differential form, and $[,]_{d^s}$ stand for a class in corresponding cohomology.
The name Gauss-Manin means that in the interpretation of connection on periods of the hypersurface this is actually a Gauss-Manin connection.

One flat connection over the contractible base contains no interesting information. However, K.Saito invented another connection that I will call K.Saito's connection.
Its definition is a bit involved and goes in several steps.

Step 1. Consider projection $\pi$ - evaluation at $z=0$:
\begin{equation}
\pi: H_{t,z} \rightarrow H_{t,0}
\end{equation}

Let us take $S$ - a section that inverts projection:
\begin{equation}
S:  H_{t,0} \rightarrow H_{t,z} ;  \; \; \;    \pi \circ S =id
\end{equation}

Step 2. Let $h \in H_{t,z}$, let $\omega$ be a $d_{t,0}^S$ closed representative of this class, $[\omega]= h$.
We will define the connection along the tangent vector $\epsilon_k$ by the result of a parallel transport along $\epsilon$ $T_{\epsilon}^S$, that acts on
$h$ as follows:

\begin{equation}
T_{\epsilon}^S (h)=
 [\omega+ z^{-1} (- \sum_{k=1}^\mu \epsilon_k \Phi_k +  S \circ \pi  (\sum_{k=1}^\mu \epsilon^k \Phi_k))\omega   ]_{d_{t+\epsilon,z}^S}
\end{equation}
The following comments are needed.
First, the term in the brackets actually is $d_{t+\epsilon,z}^S$ closed ( as can be easily seen by comparison with the
GM connection).

Second, both terms in the bracket are closed with respect to $d_{t,0}^S$, and their classes in $d_{t,0}^S$ cohomology
are equal, thus, the class of $d_{t+\epsilon,z}^S$ cohomology of bracket is vanishes at $z=0$, so the results of such
transport actually belongs to $ H_{t+\epsilon,z}$, and not just to  $\hat{H}_{t,z}$ (note, that the parallel transport in GM connection belongs to   $\hat{H}_{t+\epsilon,z}$ ).

Third, if we localize K.Saito connection at $z=0$ (i.e. consider it as connection on  $\hat{H}_{t,z}$) we can compare it
with the GM connection and we will get the following remarkable relation

\begin{equation}
\nabla^S=\nabla^{GM}+z^{-1} \sum_{k=1}^{\mu} dt_k C_k
\end{equation}

Fourth, for a general section the K.Saito connection is not integrable.
However, if the section is parallel along the K.Saito connection the K.Saito connection is integrable.
Thus, K.Saito defined a good section, such that if we find it at zero it could be transported to the full formal disk
$Spec \; \mathbb{C} [[t_1, \ldots, t_{\mu}]]   $.
 
Moreover, given a good section one can trivialize GM connection and in such trivialization K.Saito connection reads

\begin{equation}
(\nabla^S)_{GM \, frame}=d+z^{-1}A
\end{equation}

From the flatness of K.Saito connection it follow that:
\begin{equation} \label{S2}
A^2=0  \; \;
\end{equation}
\begin{equation} \label{S1}
dA=0
\end{equation}
Note, that system (\ref{S2}) and (\ref{S1}) correspond exactly to Commutativity equations!
That is why the Commutativity equations were actually discovered by K.Saito.

It looks like a miracle. In \cite{Lo} this was explained as follows.

There is a analogue of complex Hodge theory where the role of complex structure is played by function $W_t$.
The differential $Q_t$ depends on $W_t$ while differential $G_{-}$ does not. Hodge property identifies cohomology of
$Q$ and $G_{-}$ through harmonic forms that are simultaneously annihilated by $Q$ and $G_{-}$.
K.Saito connection is such a connection in cohomology of $Q_t$ that corresponding class in $G_{-}$ cohomology
is not changing. That is why it is integrable.  The good section is a class of holomorphic part of germ of harmonic form at singularity, and Hodge connection restricted to classes of $Q_t+zG_{-}$ coincides with K.Saito's connection.

The Hodge theory allows to construct the solution to Commutativity equation, that is why K.Saito theory of good section does the same.

\end{document}